\documentclass[amsmath,amssymb,aps,prx,superscriptaddress,reprint]{revtex4-2}
\usepackage{graphicx}

\begin{document}

\title{Dynamical Scaling Reveals Topological Defects and Anomalous Evolution of a Photoinduced Phase Transition} 

\author{Gal Orenstein}\email{galore@slac.stanford.edu}
\affiliation{Stanford PULSE Institute, SLAC National Accelerator Laboratory, Menlo Park, California 94025, USA}
\affiliation{Stanford Institute for Materials and Energy Sciences, SLAC National Accelerator Laboratory, Menlo Park, CA 94025, USA.}
\author{Ryan A. Duncan}
\affiliation{Stanford PULSE Institute, SLAC National Accelerator Laboratory, Menlo Park, California 94025, USA}
\affiliation{Stanford Institute for Materials and Energy Sciences, SLAC National Accelerator Laboratory, Menlo Park, CA 94025, USA.}
\author{Gilberto A. de la Pe$\tilde{\text{n}}$a Mu$\tilde{\text{n}}$oz}
\affiliation{Stanford PULSE Institute, SLAC National Accelerator Laboratory, Menlo Park, California 94025, USA}
\affiliation{Stanford Institute for Materials and Energy Sciences, SLAC National Accelerator Laboratory, Menlo Park, CA 94025, USA.}
\author{Yijing Huang}
\affiliation{Stanford PULSE Institute, SLAC National Accelerator Laboratory, Menlo Park, California 94025, USA}
\affiliation{Stanford Institute for Materials and Energy Sciences, SLAC National Accelerator Laboratory, Menlo Park, CA 94025, USA.}
\author{Viktor Krapivin}
\affiliation{Stanford PULSE Institute, SLAC National Accelerator Laboratory, Menlo Park, California 94025, USA}
\affiliation{Stanford Institute for Materials and Energy Sciences, SLAC National Accelerator Laboratory, Menlo Park, CA 94025, USA.}
\author{Quynh Le Nguyen}
\affiliation{Linac Coherent Light Source, SLAC National Accelerator Laboratory, Menlo Park, California 94025, USA}
\author{Samuel Teitelbaum}
\affiliation{Department of Physics, Arizona State University, Tempe, Arizona 85281, USA}
\author{Anisha G. Singh}
\affiliation{Department of Applied Physics, Stanford University, Stanford, California 94305, USA}
\author{Roman Mankowsky}
\affiliation{Paul Scherrer Institut, Villigen, Switzerland}
\author{Henrik Lemke}
\affiliation{Paul Scherrer Institut, Villigen, Switzerland}
\author{Mathias Sander}
\affiliation{Paul Scherrer Institut, Villigen, Switzerland}
\author{Yunpei Deng}
\affiliation{Paul Scherrer Institut, Villigen, Switzerland}
\author{Christopher Arrell}
\affiliation{Paul Scherrer Institut, Villigen, Switzerland}
\author{Ian R. Fisher}
\affiliation{Department of Applied Physics, Stanford University, Stanford, California 94305, USA}
\author{David A. Reis}
\affiliation{Stanford PULSE Institute, SLAC National Accelerator Laboratory, Menlo Park, California 94025, USA}
\affiliation{Stanford Institute for Materials and Energy Sciences, SLAC National Accelerator Laboratory, Menlo Park, CA 94025, USA.}
\author{Mariano Trigo}\email{mtrigo@slac.stanford.edu}
\affiliation{Stanford PULSE Institute, SLAC National Accelerator Laboratory, Menlo Park, California 94025, USA}
\affiliation{Stanford Institute for Materials and Energy Sciences, SLAC National Accelerator Laboratory, Menlo Park, CA 94025, USA.}



\begin{abstract}
Nonequilibrium states of quantum materials can exhibit exotic properties and enable unprecedented functionality and applications. These transient states are inherently inhomogeneous, characterized by the formation of topologically protected structures, requiring nanometer spatial resolution on femtosecond timescales to resolve their evolution. Using ultrafast total x-ray scattering at a free electron laser and a sophisticated scaling analysis, we gain unique access to the dynamics on the relevant mesoscopic lengthscales. Our results provide direct evidence that ultrafast excitation of LaTe$_3$ leads to formation of topological vortex strings of the charge density wave. These dislocations of the charge density wave exhibit anomalous, subdiffusive dynamics, slowing the equilibration process, providing rare insight into the nonequilibrium mesoscopic response in a quantum material. Our findings establish a general framework to investigate properties of topological defects, which are expected to be ubiquitous in nonequilibrium phase transitions and may arrest equilibration and enhance competing orders.
\end{abstract}
\maketitle 

\section{Introduction}
One of the defining features of complex matter is the competition between several active degrees of freedom, which leads to highly heterogeneous, often modulated ground states in quantum materials\cite{Dagotto2005Complexity}.
The relationship between these spatial fluctuations, topological defects and competing phases remains a fundamental problem in condensed matter physics \cite{Mitrovic2001Spatially,Kalisky2010stripes,Mesaros2011,Joe2014emergence,Yan2017influence,Straquadine2019suppression,Yu2019fragile,Fang2019disorder}.
Ultrafast light pulses could provide insight into this relation by driving nonequilibrium phase transitions \cite{Basov2017,schmitt2008,fausti@2011,Ichikawa2011Transient,Huber2014,Stojchevska2014,nova2019,xian2019,Sun2020transient,Kogar2020,Ravnik2021,Vogelgesang2018,Wandel2022}, however visualizing the transient structures at the relevant mesoscopic lengthscales is challenging. 
Novel x-ray free electron laser (XFEL) sources offer the sensitivity to all lengthscales and can probe the spatial complexity of these transient states on ultrafast timescales \cite{wall2018}. Here we use an XFEL to unveil the mesoscopic dynamics of topological defects of charge order following a nonequilibrium symmetry breaking phase transition. 

In a material undergoing a symmetry breaking phase transition under nonequilibrium conditions, the low-symmetry phase develops independently in separate regions of space, leaving behind disconnected domains.  This leads to the formation of topological defects (TDs) by the Kibble-Zurek mechanism \cite{Kibble_1976,Zurek1985}. 
As the system evolves towards equilibrium, these defects annihilate and the independently ordered domains coalesce. Generally, this coarsening process obeys self-similar dynamic scaling\cite{Mondello1992,Bray1994Growth,Wong1993,Vogelgesang2018,Mitrano2019}, where the fluctuations are characterized by a single lengthscale, $L(t)$. Here $L(t)$ corresponds to a mean domain size, or an average distance between TDs. 
Under such conditions, the structure factor $S\left(\mathbf{k}, t\right)$, as measured by x-ray scattering, has the form:
\begin{equation}
\begin{gathered}
    S\left(\mathbf{k},t\right)=g(t)F\left(\mathbf{k}L(t)\right)\\
    L(t)= (A t)^\beta
\end{gathered}
\end{equation}

where $g(t)$ is an amplitude which depends only on time and scales the intensity equally for all wavevectors, $F\left(\mathbf{x}\right)$ is a universal function, $A^\beta$ a proportionality constant and $\beta$ determines whether the coarsening is superdiffusive ($\beta>1/2$), diffusive ($\beta = 1/2)$ or subdiffusive ($\beta < 1/2)$. The scaling of $F\left(\mathbf{x}\right)$, as well as the value of the growth exponent, $\beta$, are powerful indicators of the symmetry of the order parameter and reveal microscopic properties and conserved quantities\cite{Bray1994Growth,hohenberg1977,Pretko2018Fracton,Gromov2020FractonHydrodynamics}. 

In spite of the long history of theoretical works \cite{Mondello1992,Mazenko1985,Zurek1985}, experimental observation of TD dynamics is notoriously difficult and remains an active area of research. Previous works have presented evidence of the central role of TDs in the evolution of quantum materials far from equilibrium\cite{Yusupov2010,Mertelj2013,Vogelgesang2018,Zong2019}. However, understanding the impact of TDs on transient photoinduced states poses intricate challenges due to the lack of suitable probes at the intrinsic nanometer length- and picosecond time-scales of the defects. To a large extent, photoinduced states of materials are still interpreted in terms of time-dependent, homogeneous order parameters, even though their recovery should behave more like a transient glass of topological defects, greatly affecting their properties.  

Here we use the unprecedented temporal and momentum resolution provided by an x-ray free electron laser to reveal the existence of photoinduced topological defects and capture their dynamics. We study the photoinduced coarsening process of the charge order in the prototypical charge density wave (CDW) material LaTe$_3$, which is a member of the rare-earth tritellurides. These materials possess a layered, quasi-two-dimensional crystal structure and undergo a second-order phase transition with temperature into a CDW state directed along the in-plane $c$ axis\cite{ru2008,yao2006theory,ru2008thesis}. Recently, the rare-earth tritellurides have attracted attention as a model system to investigate the dynamics of symmetry-breaking phase transitions \cite{schmitt2008,Yusupov2010,Zong2019,Kogar2020,Trigo2021,Wang2022}. 
The high quality data from the XFEL enables a sophisticated scaling analysis of the structure factor. This incisive analysis reveals unequivocal signatures of vortex strings, which govern the return to equilibrium with a growth exponent of $\beta = 0.29$. The subdiffusive coarsening process is likely due to restricted microscopic mobility of the CDW dislocations corresponding to the vortex strings.

\section{Experimental signatures of coarsening}
We probed the dynamics of the lattice component of the CDW in LaTe$_3$ using 50 fs, 10keV hard x-ray pulses at the Bernina instrument \cite{ingold2019} of the SwissFEL facility\cite{Prat2020}. The dynamics were initiated by a 35 fs optical pump pulse as illustrated in Fig. 1(a) (see Appendix A). The following results were taken with a pump fluence of 8 mJ/cm$^2$. Fig. 1(b) shows the detector image of the $\mathbf{G}=\left(2/a,2/b,(1-q_{CDW})/c\right)$ CDW Bragg peak, where $a$, $b$ and $c$ are the respective standardized cell parameters and $q_{CDW}\approx2/7$. The intensity is shown with logarithmic scale and integrated over pump-probe delays between $-2$ and $22$ ps. The peak elongates in the $b^*$ direction due to shorter correlations along the out-of-plane $b$ axis compared to the in-plane $a$-$c$ axes.

\begin{figure}[ht]
\centering
\includegraphics[trim=0 0 0 0,clip,width=1\linewidth]{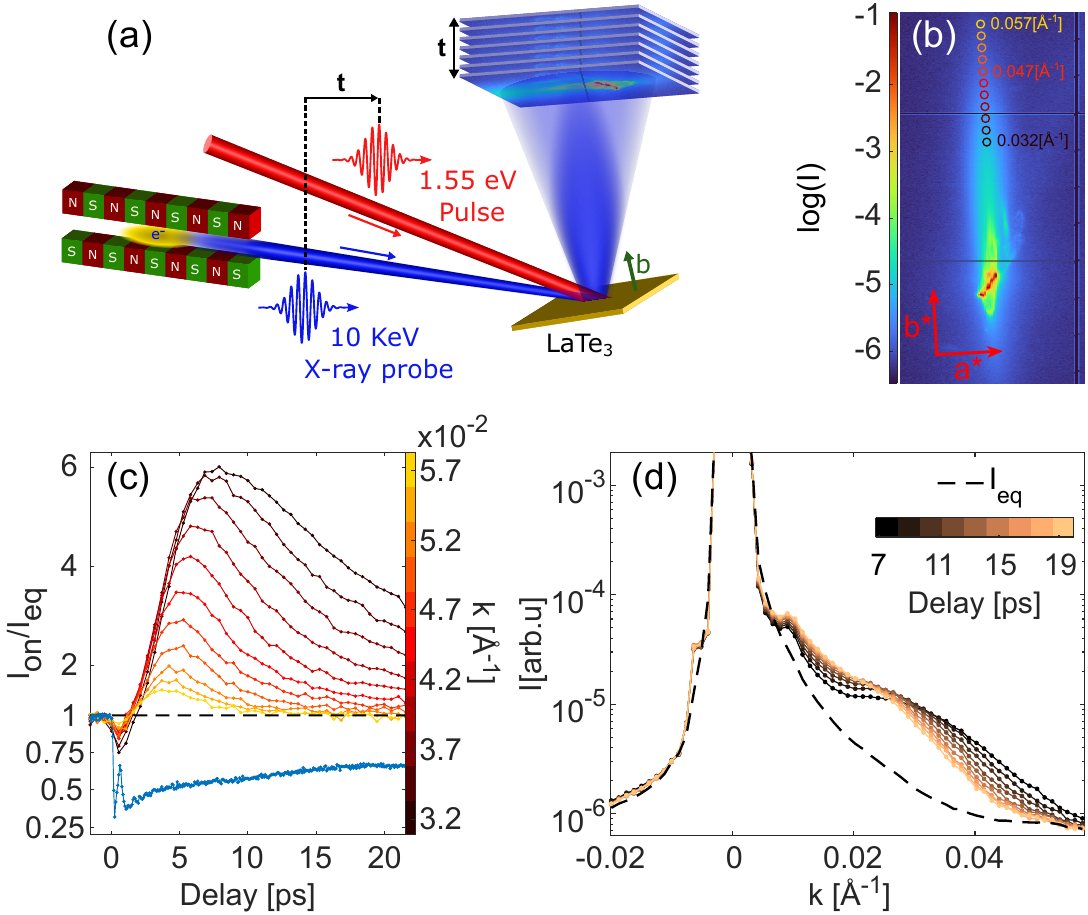}
\caption{X-ray scattering signatures of coarsening dynamics. (a), Experimental setup. The dynamics were initiated by an intense 800nm pulse. The temporal evolution of the CDW's lattice component was probed by 10keV hard x-ray pulses delayed by $t$ with respect to the 800nm pulses. (b), Detector image of the $\left(2,2,1-q_{CDW}\right)$ superlattice peak, integrated over all pump-probe delays. The red arrows indicate directions on the detector which are predominantly oriented along the $b^*$ and $a^*$ axes. The circles correspond to the black to yellow lines in (c).
(c), x-ray intensity versus $t$, normalized by the equilibrium intensity. The blue line shows the normalized intensity near the peak. 
The black to yellow lines correspond to the same coloured circles in (b).
(d), x-ray intensity on a logarithmic scale versus wavevector along $\mathbf{G}+\mathbf{k}$, for $\mathbf{k}\approx\left(0,k,0\right)$, at different pump-probe delays starting from 6.5 ps. The dashed black line is the equilibrium intensity. The intensity in (c) and (d) is averaged over a time window and regions of the detector as described in Appendix B(1).}
\end{figure}

Fig. 1(c) shows the intensity normalized by the equilibrium value as a function of pump-probe delay, $t$, at the representative regions along the $b^*$ direction marked in Fig. 1(b). The intensity near the peak (blue line), which represents the long-range order of the CDW, decreases sharply after $t = 0$ and remains depressed below the equilibrium level throughout our measurement. Generally, photoexcitation above the CDW bandgap induces a fast suppression of the long-range order, which subsequently recovers with an extremely long timescale depending on the strength of the excitation\cite{Mertelj2013,Huber2014,Vogelgesang2018,Trigo2021}. In contrast, on the tail of the peak (black to yellow lines), the initial decrease in normalized intensity due to the suppression of the CDW modulation is followed by an increase well above equilibrium values over several picoseconds. This increase starts after $t \approx 1$~ps, which is roughly the time it takes for the CDW electronic gap to restore according to time and angle resolved photo-electron spectroscopy measurements \cite{Zong2019,schmitt2008}. The increase of diffuse intensity away from the CDW wavevector is a signature of growing inhomogeneity and spatial fluctuations of the CDW induced by photoexcitation. Its observation is enabled by the high brightness of the XFEL together with its exceptional temporal and wavevector resolution, which allow subtle details of the dynamics of the diffuse intensity on the tails of the peak to be captured.

In Fig. 1(d) we show the scattering intensity, $I(k,t)$, as a function of wavevector along $\mathbf{G}+\mathbf{k}$, for $\mathbf{k}\approx\left(0,k,0\right)$, at representative delays after $\sim 6.5$ ps. Comparing these plots with the equilibrium intensity shown by the black dashed line, it is clear that the peak broadens initially (black dots) and gradually narrows back at later times (copper dots). The narrowing of the peak is the hallmark of coarsening of spatially inhomogeneous regions as the system recovers towards equilibrium.

\section{Observation of subdiffusive dynamics}
To extract the growth exponent of the coarsening process, $\beta$ in Eq. (1), it is convenient to define $h\left(k,t\right)=k^2 S\left(k,t\right)$. In Fig. 2(a) we show the measured $h\left(k,t\right)$ by plotting $k^2I(k,t)$ at several representative delays, normalizing each trace by its maximal value. The traces exhibit a distinct maximum which visibly shifts horizontally to lower $k$ values with increasing $t$. Since $h(k, t)$ has the same scaling form as $S(k, t)$, the $k$ position of the maximum as a function of time, $k_{m}(t)$, satisfies $k_{m}(t)\propto L^{-1}(t)\propto t^{-\beta}$ because $L(t)$ is the only relevant lengthscale, completely governing the system's evolution. Furthermore, $k_{m}(t)$ is independent of $g(t)$ because $g(t)$ scales the intensity equally for all wavevectors (see Appendix B(2)). To robustly resolve the horizontal shift of the maximum shown in Fig. 2(a), we fit the maximum region for each delay to a Gaussian and identify $k_{m}(t)$ with the Gaussian's center. In Fig. 2(b) we plot $\log(k_{m}(t))$ versus $\log(t)$. The plot fits a straight line very well with a slope of $\beta=0.29$. In the supplemental material Fig. S1 we plot similar traces to Fig. 2 for fluences of 4 mJ/cm$^2$, 6 mJ/cm$^2$, and 9 mJ/cm$^2$, showing no notable change to $\beta$. This demonstrates that the power-law is an intrinsic property of the system, which does not depend on the excitation strength. In Fig. 2(c), we plot the same traces as in Fig. 2(a), with the horizontal axis re-scaled for each delay by $L(t) \propto t^\beta$ with $\beta=0.29$. The maxima, as well as the entire region with $kL(t)>0.02$ collapse quite well for all presented delays. The deviation from scaling at low $k$ is expected since scaling only applies at lengthscales smaller than the system size. In the photoexcited system, the system size is effectively the pumped volume, which is limited by the laser penetration depth.  Here, the stationary local maximum around $k=0.01${\AA} in Fig. 2(a) is associated with the 100\AA-200{\AA} penetration depth of the laser \cite{Trigo2021}, so a robust data collapse is observed up to lengthscales of $\sim$50\AA. This data collapse clearly shows that the system exhibits a coarsening lengthscale $L(t)\propto t^{0.29}$.

\begin{figure}[ht]
\centering
\includegraphics[trim=0 0 0 0,clip,width=1\linewidth]{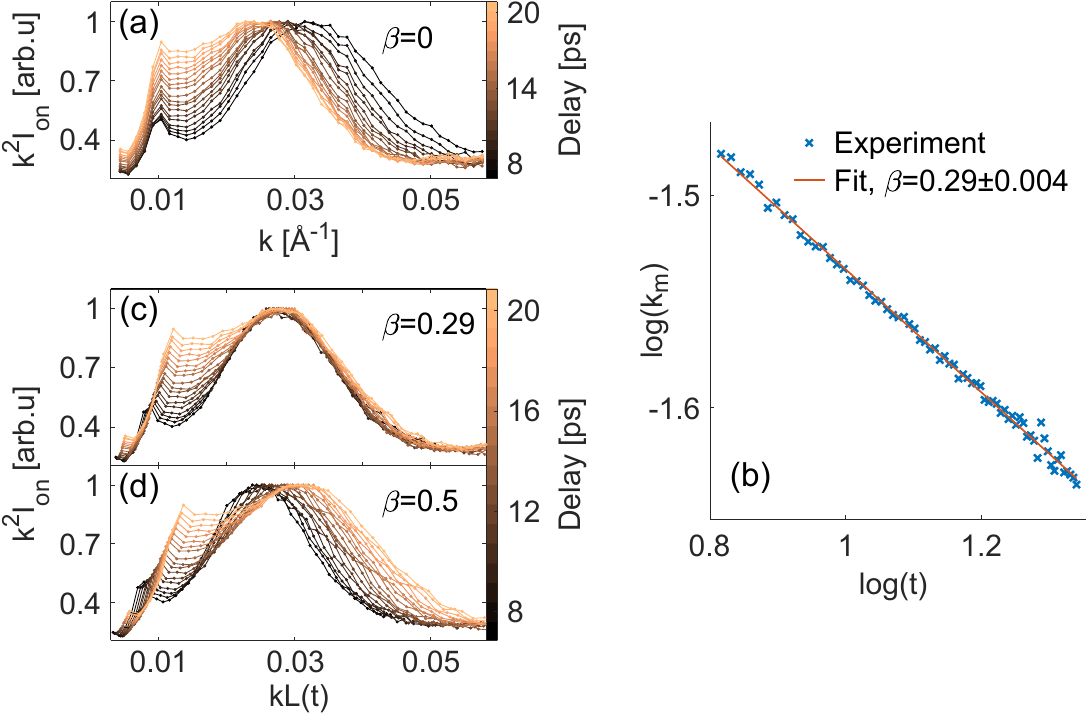}
\caption{Dynamic scaling exhibiting a subdiffusive growth exponent. (a), $k^2I(k,t)$ (the measured $h\left(k,t\right)$) plotted at different delays. The vertical axis was normalized by the maximum value for each delay. (b), $\log (k_{m}(t))$ versus $\log(t)$. The red solid straight line is the fit to the data with $\beta=0.29$. (c) and (d), the traces plotted in (a) plotted with the horizontal axis scaled by $L(t)=(A t)^\beta$ per delay for $\beta=0.29$ and 0.5 respectively. In both (c) and (d) $A$ was chosen to be 1/12 \AA$^{\frac{1}{\beta}}$/ps for better visualization. Note that the traces in (a) correspond to $\beta=0$.
 } 
\end{figure}

A growth exponent of $\beta=0.29$ is an intriguing result. Extensive theoretical studies\cite{Mazenko1985,Bray1994Growth} concluded that, in systems with a non-conserved order parameter, the recovery of long range order at long times follows a diffusive process with $\beta = 0.5$. This has been confirmed experimentally in diverse systems from nematic liquid crystals\cite{Wong1993} to CDWs\cite{Vogelgesang2018}. In Fig. 2(d) we re-scale the horizontal axis of each trace in Fig. 2(a) by $L(t) \propto t^\beta$ for $\beta=0.5$. The data collapses visibly better for $\beta=0.29$ (Fig. 2(c)) than for $\beta=0.5$ (Fig. 2(d)), confirming that the system indeed behaves subdiffusively. Subdiffusive behaviour suggests that the dynamics are constrained by additional conserved quantities. It has been established that for a system with a conserved order parameter $\beta = 0.25$ \cite{Mazenko1985}. While the appropriate order parameter, which theoretically describes the CDW here, is quite different from a conserved order parameter, the anomalous subdiffusive dynamics is a manifestation of restricted microscopic motion of the system. 

\section{Scattering signatures of vortex strings}

Our following analysis shows clear evidence that the coarsening process is dominated by photoinduced vortex strings, which are dislocations of the CDW. The subdiffusive behaviour is likely a signature of the restricted mobility of these dislocations. At large wavevectors, where interference between multiple TDs is negligible, the structure factor for a network of TDs has the asymptotic form of an isolated defect. This asymptotic form is universal and depends only on $n$, the number of components of the vector order parameter and $d$, the spatial dimension of the system\cite{Bray1993}:
\begin{equation}
    S_{TD}\left(k,t\right)\propto\rho(t)k^{-(n+d)}
\end{equation}
where $\rho(t)$ is the density of TDs, which under reasonable conditions also exhibits power-law behaviour, $\rho(t)\propto t^{\alpha}$ \cite{Mondello1992,Bray1993}. For the incommensurate CDW considered here the expectation is that $n=2$ and $d=3$, thus the stable TDs are vortex strings \cite{Mondello1992} with the asymptotic structure factor $S(k,t) \sim \rho(t)k^{-5}$. 

The unique properties of the XFEL enable us to measure the subtle intensity changes around the CDW peaks with high sensitivity and wavevector resolution and search for these power-law signatures of the TDs [Eq.(2)]. The black to yellow dotted lines in Fig. 3(a) show the scattering intensity for $k>0.02$\AA$^{-1}$ at different times in a log-log plot and the blue line is a $k^{-5}$ power-law. Clearly, at large $k$, the traces for different times, as represented by the solid lines with markers, exhibit a similar $k^{-5}$ behaviour. To extract the scaling we take the 174 data points with markers in Fig. 3(a) and fit them to $Ct^{\alpha}k^{\eta}$ extracting $\alpha=-0.96$ and $\eta=-4.96$. The red and blue symbols in the log-log plot in Fig. 3(b) show $I(k,t)t^{-\alpha}$ and $I(k,t)k^{-\eta}$, which are the data points collapsed in $t$ and $k$ respectively, using the extracted exponents $\alpha$ and $\eta$. This remarkable data collapse indicates that the asymptotic scattering is governed by Eq. (2) with $n+d=-\eta=4.96$. In supplemental material Fig. S2 we show the data collapse in $t$ for fluences of 4 mJ/cm$^2$, 6 mJ/cm$^2$, and 9 mJ/cm$^2$, resulting in similar values for $\eta$. The fact that $n+d\approx5$ independent of the pump fluence, provides decisive evidence that vortex strings dominate the scattering in the high $k$ tail following the ultrafast quench. The analysis presented here has been carried out along the out-of-plane direction, however, the exponent $\eta$ and the coarsening dynamics are independent of the anisotropy in this system (see supplemental material).

\begin{figure}[ht]
\centering
\includegraphics[trim= 0 0 0 0,clip,width=1\linewidth]{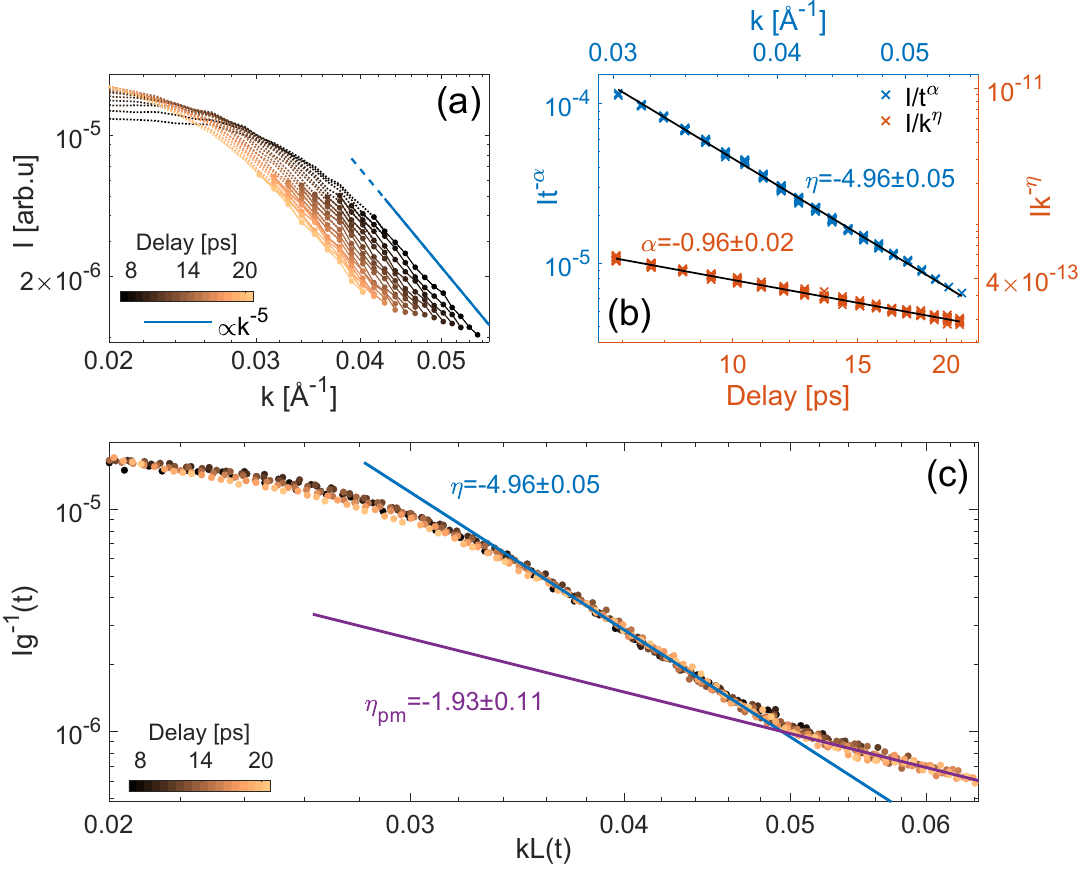}
\caption{Data collapse and vortex strings power-law in momentum space. (a), Scattering intensity versus $k$ for $k>0.02$\AA$^{-1}$ on logarithmic scales for different delays. The solid lines with markers highlight the power-law region at high $k$. (b), Data collapse of the 174 points marked in (a). The blue and red symbols show $I(k,t)t^{-\alpha}$ ($\alpha=-0.96$, left axis) and $I(k,t)k^{-\eta}$ ($\eta=-4.96$, right axis) versus $k$ (top) and $t$ (bottom) axes, respectively; all axes are logarithmic. The black solid straight lines are the fit to data with $\eta=-4.96$ and $\alpha=-0.96$. (c), $I(k,t)g^{-1}(t)$ versus $kL(t)$, for $kL(t)>0.02$, using only $\alpha$, $\beta$ and $\eta$; both axes are logarithmic. The different colours represent different delays and the solid straight lines show power-laws associated with different regions. }
\end{figure}

Having found a clear signature of vortex strings in the scattering, we now combine the scaling expression in Eq. (1) with the asymptotic scattering in Eq. (2) to scale the data. We find that, for the vortex string power-law region, $g(t)\propto t^{\alpha-\eta\beta}$ (see Appendix B(3)). We use the $\beta$, $\alpha$ and $\eta$ extracted from our data to determine $g(t)\propto t^{0.48}$ and, together with $L(t)$, scale $I(k,t)$ according to Eq. (1), as shown in Fig. 3(c). This plot shows remarkable self-similarity for $kL(t)>0.02$. The data scaling relies on the vortex string parameters, $\alpha$ and $\eta$, which were extracted independently of $\beta$ , yet these parameters collapse the data at lower $k$ as well, beyond the power-law region where $I(k,t) \propto t^{\alpha}k^{\eta}$. This relation between the TD scaling in $k$ and the global self-similar behaviour in time, reveals the strong connection between the vortex strings and the subdiffusive coarsening process. 

Fig. 3(c) provides additional insight into the evolution of the nonequilibrium state. Our high resolution measurements clearly identify a crossover from $\eta=-4.96$ (solid blue line) to $\eta_{pm}=-1.93$ (solid purple line) at large $kL(t)$. This crossover means that at later times, as the TD density decreases, the scattering at high $k$ is no longer dominated by the TDs; instead, scattering comes primarily from phase-modes, which have the asymptotic form $S\left(k,t\right)\sim k^{-2}$ \cite{Mazenko1985}. 

\section{Phenomenological Ginzburg-Landau model}

To visualize the vortex string dynamics we simulated the Ginzburg-Landau model \cite{Mondello1992} for a complex order parameter ($n=2$), $\psi(\mathbf{r}, t)$, in three spatial dimensions ($d=3$) (see Appendix C and supplemental material movie 1). Figs. 4(a)-(c) show the surface $|\psi(\mathbf{r}, t)|^2 = 0.5$ following a sudden quench to the low symmetry potential energy, at representative delays of 4 ps, 8 ps and 22 ps respectively. The initial random configuration of the disordered state around $\psi(\mathbf{r}, t)=0$ quickly breaks down into strings whose density decreases with time\cite{Mondello1992}. Fig. 4(d) shows the phase of $\psi(\mathbf{r}, t)$ winding by $2\pi$ around the string center on the plane $z=20$ indicated in Fig. 4(c); this winding of the phase occurs along the entire string length and is a topological invariant. Therefore, vortex strings can only vanish by annihilating with other vortex strings with an opposite winding number, or by shrinking of closed loops \cite{Mondello1992}. The solid symbols in Fig. 4(e) show the simulated $S(k, t)$ versus $k$ on a log-log plot and the solid lines are power-law fits. The blue symbols exhibit an asymptotic power-law with an exponent of $\eta = -5.2$, which agrees well with our experimental observations (Fig. 3). In contrast, the green symbols in Fig. 4(e) correspond to a thermal equilibrium starting from the ordered state, $\psi(\mathbf{r}, t) = 1$. In this case the system does not produce vortex strings and it exhibits a power-law with $\eta = -1.8$, close to the theoretical $\eta = -2$ expected for phase modes. Furthermore, the blue symbols in Fig. 4(e) show a crossover between scattering dominated by vortex strings, $S \sim k^{-5}$, to scattering dominated by phase modes, $S \sim k^{-2}$. We observe a similar crossover experimentally in Fig. 3(c).

\begin{figure*}[ht]
\centering
\includegraphics[width=1\linewidth]{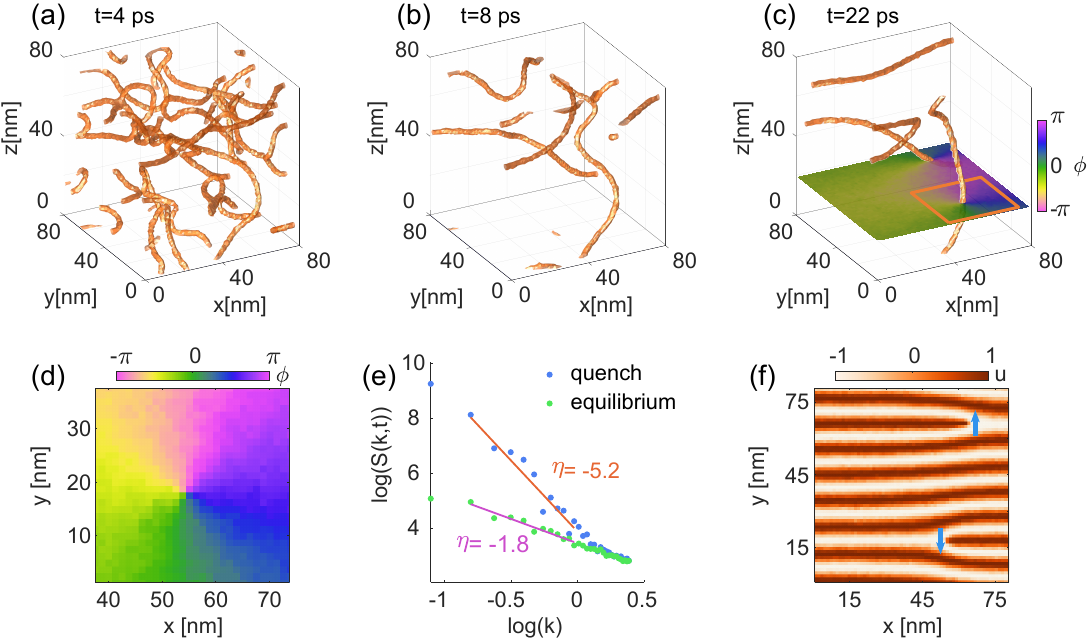}
\caption{ Ginzburg-Landau simulations of vortex strings. (a)-(c), Isosurface plots of $|\psi(\mathbf{r}, t)|^2=0.5$ at representative times after a sudden quench to the low symmetry Ginzburg-Landau potential. These plots are snapshots of the evolution shown in supplemental material movie 1. (d), a zoomed view of the order parameter phase, $\phi(\mathbf{r}, t) = \arg\{\psi(\mathbf{r}, t)\}$, on the plane indicated in (c), which winds by $2\pi$ around the vortex string. (e), Simulated structure factor, $S(k, t)$, on a log-log plot (Solid symbols), averaged between $\sim 2$ to $8$~ps. Blue dots represent $S(k, t)$ when the system is quenched suddenly from the high symmetry phase. Green symbols represent $S(k, t)$ without vortex strings (only phase modes) and was computed from thermal equilibrium starting in the ordered phase. (f), illustration of the CDW lattice distortion given by $u(x,t)\propto\cos(\mathbf{q}\cdot \mathbf{r} + \phi(\mathbf{r}, t))$ at $t=22$ ps with the $\phi(\mathbf{r}, t)$ from the plane indicated in (c). For better visibility we use $\mathbf{q}\approx\mathbf{q}_{CDW}/6$. }
\end{figure*}

In the present context, vortex strings correspond to dislocations of the CDW. Fig. 4(f) illustrates a CDW lattice distortion in the $z=20$~nm plane, taking the phase from Fig. 4(c). The vortices with opposite winding numbers correspond to two dislocations with opposite Burgers vectors \cite{nabarro1967}, indicated by blue arrows in Fig. 4(f). 
While the simple Ginzburg-Landau theory correctly predicts the formation of vortex strings with $n + d = 5$\cite{Bray1994Growth,Mondello1992},
it incorrectly predicts purely diffusive coarsening with $\beta = 0.5$\cite{Bray1994Growth,Mondello1992}. The observed subdiffusive dynamics requires a description beyond this phenomenological model, as it does not account for the anisotropic motion of the dislocations along and perpendicular to the Burgers vectors. Dislocations with restricted mobility are analogous to other systems with restricted motion, which exhibit subdiffusive thermalization and broken ergodicity due to additional conserved microscopic quantities\cite{Pretko2018Fracton,Gromov2020FractonHydrodynamics}. Clearly, our measurements and incisive analysis reveal subtle microscopic details that challenge the canonical framework used in the interpretation of many ultrafast condensed-matter experiments \cite{Yusupov2010,Huber2014,Kogar2020,Zong2019}.

\section{Conclusions}

Our x-ray diffuse scattering measurements reveal explicit hallmarks of photoinduced topological defects of the prototypical CDW in LaTe$_3$ on ultrafast time scales. We show that shortly after photoexcitation the evolution of the highly nonequilibrium state is subdiffusive, indicating that the dynamics is dominated by the restricted motion of the topological vortex strings, which crucially impacts the system's thermalization and broken ergodicity. Our methods, as well as the implications of our findings, extend well beyond LaTe$_3$, as formation of topological defects is expected to be ubiquitous in symmetry breaking phase transitions under nonequilibrium conditions \cite{Kibble_1976,Zurek1985}. Thus, the high resolution XFEL measurements, combined with scaling and data collapse analysis, establish an incisive approach to probe the mesoscopic, out-of-equilibrium behaviour of intertwined orders in quantum materials \cite{Ravnik2021}.

\section*{Acknowledgments}
We acknowledge the Paul Scherrer Institute, Villigen, Switzerland for provision of free-electron laser beamtime at the Bernina instrument of the SwissFEL ARAMIS branch. 

G.O., R.A.D, G.A.P.M., Y.H., V.K., D.A.R., and M.T. were supported by the US Department of Energy, Office of
Science, Office of Basic Energy Sciences through the Division of Materials Sciences and Engineering under Contract No. DE-AC02-76SF00515. For crystal growth and characterization, A.G.S. and I.R.F. were supported by the Department of Energy, Office of Basic Energy Sciences, under contract DE-AC02-76SF00515. G.O. acknowledges support from the Koret Foundation. R.A.D. acknowledges support through the Bloch Postdoctoral Fellowship in Quantum Science and Engineering from the Stanford University Quantum Fundamentals, Architectures, and Machines initiative (Q-FARM), and the Marvin Chodorow Postdoctoral Fellowship from the Stanford University Department of Applied Physics. Q.L.N. acknowledges support by the Q-FARM Bloch Fellowship and U.S. DOE (DE-AC02–76SF00515).

\section*{Appendix A: Experimental details}
 Single crystals of LaTe$_3$ were grown from a self flux \cite{ru2006,ru2008}. Below the transition temperature, which is estimated at $\sim670$~K \cite{hu2014}, LaTe$_3$ develops an incommensurate CDW with wavevector along the in-plane $c$ axis. The in-plane $a$ and $c$ axes of the samples were determined with x-ray diffraction by comparing the intensity of the (0,6,1) peak with its forbidden counterpart (1,6,0). The crystal was cleaved before the experiment and kept under Nitrogen gas flow to avoid oxidation. The sample normal was along the $b$ axis.
 
 Measurements were performed at the Bernina instrument \cite{ingold2019} of the SwissFEL facility \cite{Prat2020}. We probed the dynamics of the lattice component of the CDW using 50 fs monochromatized hard x-ray pulses, tuned to 10keV. The dynamics were initiated by a 35 fs optical pump pulse with a wavelength of 800 nm. Measurements were conducted in a grazing incidence geometry to achieve similar x-ray and optical penetration. The x-ray beam was incident at $1^\circ$ with respect to the sample surface, focused to a 0.01 mm $\times$ 0.2 mm spot, while the laser was impinging on the sample at an angle of $5^\circ$, focused to 0.3 mm $\times$ 0.4 mm. The scattered photons were recorded by a Jungfrau area detector positioned 482 mm from the sample. We present results for a fluence of 8 mJ/cm$^2$, however measurements were taken with 4 mJ/cm$^2$, 6 mJ/cm$^2$ and 9.7 mJ/cm$^2$ as well, showing no notable change to the exponents $\beta$ and $n+d$. 

\section*{Appendix B: Data Analysis details}
\renewcommand\thesubsection{\arabic{subsection}}
 \subsection{Averaging for Figs. 1(c) and 1(d)}
\begin{itemize}
    \item \textit{Fig. 1(c), black to yellow lines.} A 20 pixel $\times$ 20 pixel average and a 525 fs time average, corresponding to $\sim$2400 x-ray pulses per time point.
    \item \textit{Fig. 1(c), Blue Line.} A 10 pixel vertical $\times$ 6 pixel horizontal average (directions are relative to Fig. 1(b)) and a 75 fs time average, corresponding to $\sim$350 x-ray pulses per time point.
    \item \textit{Fig. 1(d)} A 10 pixel vertical $\times$ 20 pixel horizontal average (directions are relative to Fig. 1(b)) and a 828 fs time average, corresponding to $\sim$3800 x-ray pulses per time point.
\end{itemize}

\subsection{Relation between $k_m(t)$, $L(t)$ and $g(t)$}
In the main text we defined $h\left(k,t\right)=k^2 S\left(k,t\right)$ which has the same scaling form as $S\left(k,t\right)$ (Eq. (1)) with $\tilde{g}(t)=g(t)L^{-2}(t)$ and $\tilde{F}\left[kL(t)\right]=[kL(t)]^2 F\left[kL(t)\right]$. We want to show that the position of the maximum of $h\left(k,t\right)$ as a function of time satisfies the relation $k_{m}(t)=x_{m} L^{-1}(t)$, where $x_{m}$ does not depend on time or $g(t)$. The maximum of $h\left(k,t\right)$ obeys:
\begin{equation}
\begin{gathered}
     \frac{\partial h\left(k_{m}(t),t\right)}{\partial k}=g(t)\{2 k_{m}(t)F\left(k_{m}(t)L(t)\right)+\\
     k_{m}^2(t) L(t)F'\left(k_{m}(t)L(t)\right)\}=0
\end{gathered}
\end{equation}
where $F'(x) = d F(x)/dx$. Clearly, $g(t)$ drops out of this equation, which yields:
\begin{equation} 
    2 F(k_{m}(t)L(t))+k_{m}(t)L(t)F'(k_{m}(t)L(t))=0
\end{equation}
Eq. (5) depends on time only through the function $k_{m}(t)L(t)$, therefore it can be solved by the time-independent variable $x_m$ defined as:
\begin{equation}
   k_{m}(t)=x_{m}L^{-1}(t) 
\end{equation}
Thus, $k_{m}(t)$ has the same scaling as $L^{-1}(t)$. When $L(t)$ is a power-law, the exponent can be extracted directly from the position of the maximum, $k_{m}(t)$, as shown in the logarithmic plot in Fig. 2(b).

\subsection{Scaling of the Data, Using $\alpha$, $\beta$ and $\eta$}
To scale the measured intensity, $I\left(k,t\right)$, we assume both scaling forms Eq. (1) and Eq. (2) and find $g(t)$: 
\begin{equation}
\begin{gathered}
     S\left(k,t\right)= Ct^{\alpha-\eta\beta}(kL(t))^{\eta}\\
     g(t)=Ct^{\alpha-\eta\beta}.
\end{gathered}
\end{equation}

 \section*{Appendix C: Ginzburg-Landau Simulation details}
We performed Ginzburg-Landau simulations in three spatial dimensions for a complex order parameter including a random Langevin force. The temporal evolution was simulated using the Euler–Maruyama method and the spatial derivatives were calculated with a Fourier Spectral Method. The order parameter, $\psi(\mathbf{r}, t)$, in the low symmetry free energy evolves according to \cite{Mondello1992}:
\begin{equation}
    \Gamma \frac{\partial \psi}{\partial t} = \{1 - |\psi|^2\}\psi + \xi^2\nabla^2 \psi + \zeta
\end{equation}
where $\Gamma=4.8$ ps is a phenomenological damping, $\xi = 1.2$~nm is the coherence length of the order parameter and $\zeta$ is the random Langevin force. For simplicity we have assumed the terms to be isotropic, which should not affect the general conclusion extracted from the simulations. The simulations were carried-out on a 64$\times$64$\times$64 grid representing a volume of 80 nm $\times$ 80 nm $\times$ 80 nm, with a time step of 20.8 fs over a range of 25 ps.

For the vortex string results (Figs. 4(a)-(d) and blue symbols in (e)) we started with a random initial condition with zero spatial mean, $\int d^3r\,\psi(\mathbf{r}, 0) = 0$. This corresponds to the system suddenly being quenched to the low symmetry free energy at $t>0$, forming TDs. The thermal equilibrium simulation (green symbols in Fig. 4(e)) started from an ordered state of $\psi(\mathbf{r}, 0)=1$ and only develops low energy excitations i.e. phase modes.

The structure factor is given by $S(k, t)=|\tilde{\psi}(\mathbf{r}, t)|^2$, where $\tilde{\psi}(\mathbf{k}, t)$ is the Fourier transform of $\psi(\mathbf{r}, t)$. The graphs in Fig. 4(e) were obtained by averaging $S(k, t)$ between 2 ps to 8 ps.

%

\end{document}


\title{Supplementary Information} 

\author{Gal Orenstein}\email{galore@slac.stanford.edu}
\affiliation{Stanford PULSE Institute, SLAC National Accelerator Laboratory, Menlo Park, California 94025, USA}
\affiliation{Stanford Institute for Materials and Energy Sciences, SLAC National Accelerator Laboratory, Menlo Park, CA 94025, USA.}
\author{Ryan A. Duncan}
\affiliation{Stanford PULSE Institute, SLAC National Accelerator Laboratory, Menlo Park, California 94025, USA}
\affiliation{Stanford Institute for Materials and Energy Sciences, SLAC National Accelerator Laboratory, Menlo Park, CA 94025, USA.}
\author{Gilberto A. de la Pe$\tilde{\text{n}}$a Mu$\tilde{\text{n}}$oz}
\affiliation{Stanford PULSE Institute, SLAC National Accelerator Laboratory, Menlo Park, California 94025, USA}
\affiliation{Stanford Institute for Materials and Energy Sciences, SLAC National Accelerator Laboratory, Menlo Park, CA 94025, USA.}
\author{Yijing Huang}
\affiliation{Stanford PULSE Institute, SLAC National Accelerator Laboratory, Menlo Park, California 94025, USA}
\affiliation{Stanford Institute for Materials and Energy Sciences, SLAC National Accelerator Laboratory, Menlo Park, CA 94025, USA.}
\author{Viktor Krapivin}
\affiliation{Stanford PULSE Institute, SLAC National Accelerator Laboratory, Menlo Park, California 94025, USA}
\affiliation{Stanford Institute for Materials and Energy Sciences, SLAC National Accelerator Laboratory, Menlo Park, CA 94025, USA.}
\author{Quynh Le Nguyen}
\affiliation{Linac Coherent Light Source, SLAC National Accelerator Laboratory, Menlo Park, California 94025, USA}
\author{Samuel Teitelbaum}
\affiliation{Department of Physics, Arizona State University, Tempe, Arizona 85281, USA}
\author{Anisha G. Singh}
\affiliation{Department of Applied Physics, Stanford University, Stanford, California 94305, USA}
\author{Roman Mankowsky}
\affiliation{Paul Scherrer Institut, Villigen, Switzerland}
\author{Henrik Lemke}
\affiliation{Paul Scherrer Institut, Villigen, Switzerland}
\author{Mathias Sander}
\affiliation{Paul Scherrer Institut, Villigen, Switzerland}
\author{Yunpei Deng}
\affiliation{Paul Scherrer Institut, Villigen, Switzerland}
\author{Christopher Arrell}
\affiliation{Paul Scherrer Institut, Villigen, Switzerland}
\author{Ian R. Fisher}
\affiliation{Department of Applied Physics, Stanford University, Stanford, California 94305, USA}
\author{David A. Reis}
\affiliation{Stanford PULSE Institute, SLAC National Accelerator Laboratory, Menlo Park, California 94025, USA}
\affiliation{Stanford Institute for Materials and Energy Sciences, SLAC National Accelerator Laboratory, Menlo Park, CA 94025, USA.}
\author{Mariano Trigo}\email{mtrigo@slac.stanford.edu}
\affiliation{Stanford PULSE Institute, SLAC National Accelerator Laboratory, Menlo Park, California 94025, USA}
\affiliation{Stanford Institute for Materials and Energy Sciences, SLAC National Accelerator Laboratory, Menlo Park, CA 94025, USA.}

\maketitle
\newpage
\section*{Fluence Dependence}
 \begin{figure}[ht]
\centering
\includegraphics[trim= 0 0 0 0,clip,width=0.7\linewidth]{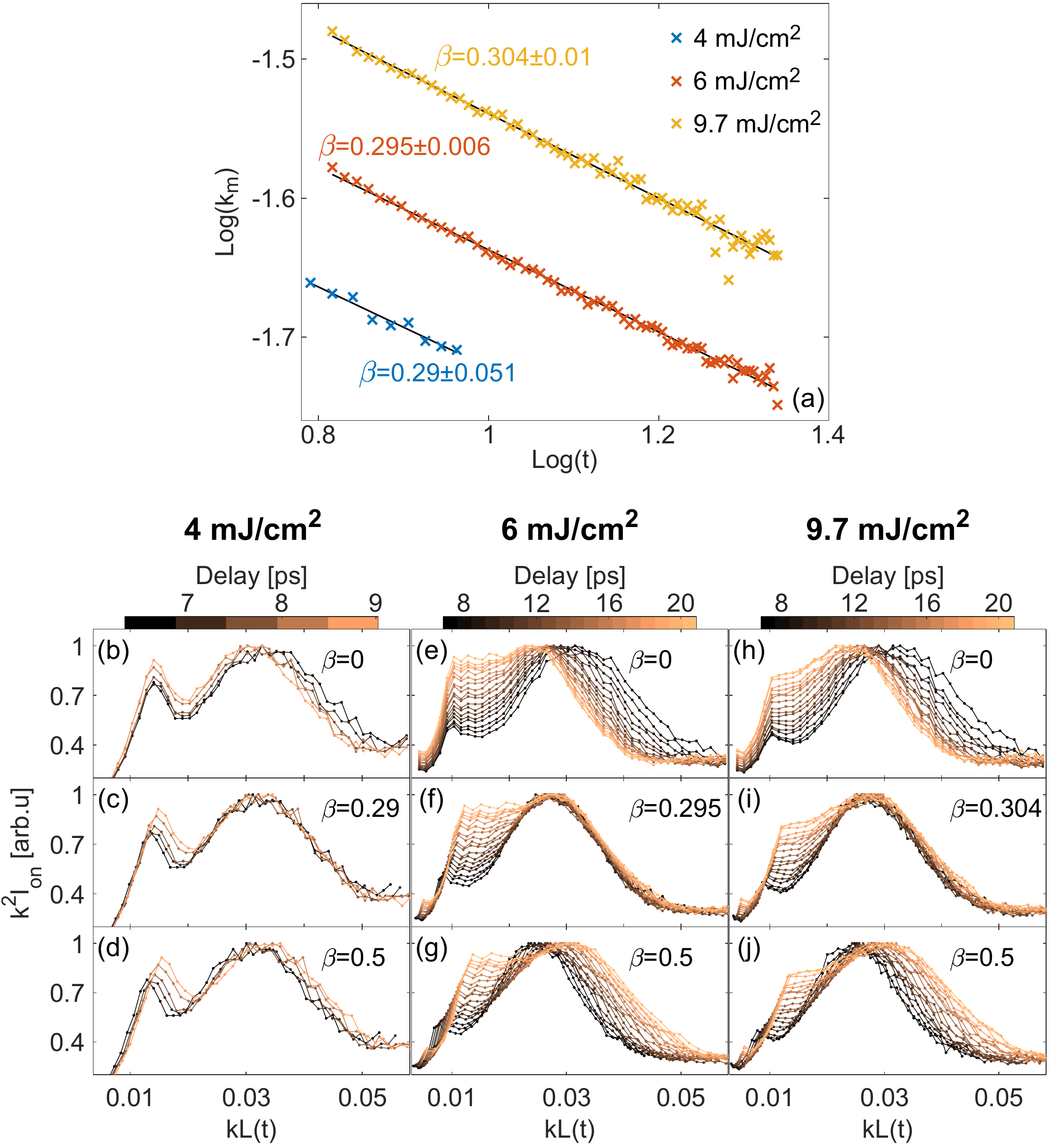}
\caption{Fluence dependence of $\boldsymbol{\beta}$. (a) $\log (k_m(t))$ versus $\log(t)$, similar to main text Fig. 2(b), showing the extracted $\beta$ for fluences of 4 mJ/cm$^2$, 6 mJ/cm$^2$ and 9.7 mJ/cm$^2$. The black straight lines are fits to the data. The data for different fluences were vertically shifted for better visibility.  (b)-(d), (e)-(g) and (h)-(j) show similar analysis to main text Figs. 2(a),(c),(d) for 4 mJ/cm$^2$, 6 mJ/cm$^2$ and 9.7 mJ/cm$^2$ respectively. The 4 mJ/cm$^2$ measurements were only taken up to 10 ps during the experiment so we chose  $A=1/7.5$ \AA$^{\frac{1}{\beta}}$/ps in (b)-(d) ($L(t)=(A t)^\beta$) for better visualization of the data at this fluence.}
\label{beta_fluence_dependence}
\end{figure}

\clearpage

 \begin{figure}[ht]
\centering
\includegraphics[trim= 0 0 0 0,clip,width=0.6\linewidth]{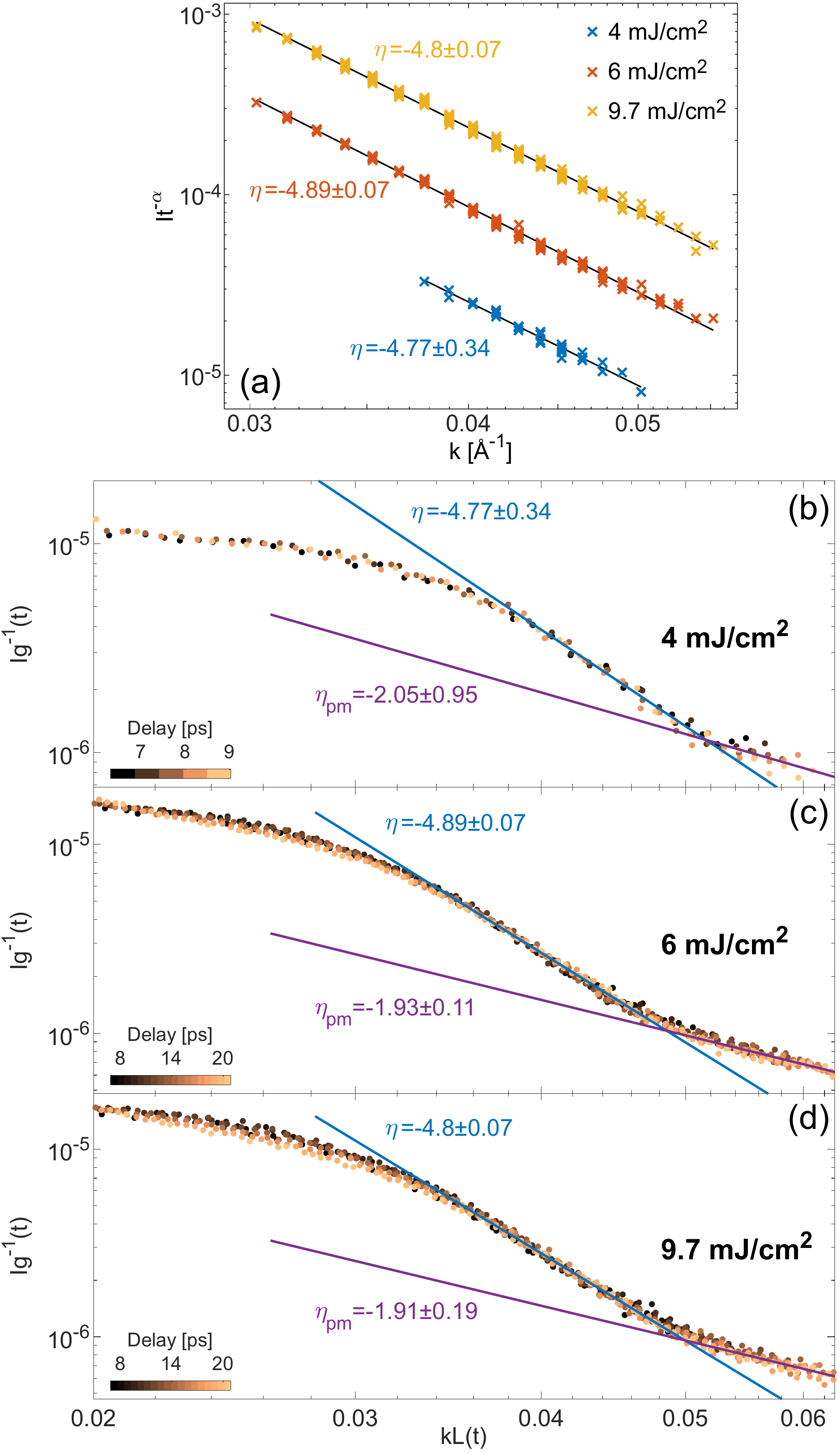}
\caption{\textbf{Fluence dependence of $\boldsymbol{\eta}$.} (a) $I(k,t)t^{-\alpha}$ versus $k$ for the points in the asymptotic power-law region on logarithmic axes, similar to main text Fig. 3(b), showing the extracted $\eta$ for fluences of 4 mJ/cm$^2$, 6 mJ/cm$^2$ and 9.7 mJ/cm$^2$. The black straight lines are fits to the data. The data for different fluences were vertically shifted for better visibility. (b)-(d) show similar analysis to main text Fig. 3(c) for 4 mJ/cm$^2$, 6 mJ/cm$^2$ and 9.7 mJ/cm$^2$ respectively. The 4 mJ/cm$^2$ measurements were only taken up to 10 ps during the experiment so we chose  $A=1/7.5$ \AA$^{\frac{1}{\beta}}$/ps in (b) ($L(t)=(A t)^\beta$) for better visualization of the data at this fluence.}
\label{eta_fluence_dependence}
\end{figure}

\clearpage

\section*{Effects of anisotropy on the coarsening and scaling}

The scattered intensity is described by the structure factor $S(\mathbf{k},t)$, which is the Fourier transform of the pair correlation function $C(\mathbf{r},t)$ \cite{Bray1994Growth,Bray1993,Mazenko1985}. Because the crystal is not cubic, the anisotropy in the correlation lengths along $a$, $b$ and $c$ means that the correlation function $C(\mathbf{r},t)$ is different along each crystal direction. However, as we show here, the exponent $\eta=-(n+d)$, is an intrinsic property that is not affected by the anisotropy in $C(\mathbf{r},t)$.

For simplicity we assume that the in-plane correlation lengths along $a$ and $c$ are the same and are larger than the out-of-plane correlation length. In the self-similar region, where the dynamics is described by a single lengthscale, we can apply a coordinate transformation, $\mathbf{r}'=\left(x',y',z'\right)=\left(x,\chi y,z\right)$, which rescales the $b$ axis by $\chi>1$, resulting in an isotropic correlation function $\tilde{C}(\mathbf{r'},t)$. In the $\mathbf{r'}$ coordinates the structure factor yields the known result for the isotropic system, $\tilde{S}_{TD}(\mathbf{k}',t)\propto\rho(t)|\mathbf{k}'|^{-(n+d)}$\cite{Bray1993}. Transforming back to the unprimed crystal coordinates gives the expression,
\begin{equation}
    S_{TD}(\mathbf{k},t)\propto\rho(t)\left(k_{x}^{2}+(k_{y}/\chi)^{2}+k_{z}^{2}\right)^{-\frac{1}{2}(n+d)}
\end{equation}

 Setting $\mathbf{k}=\mathbf{k}_{\perp}=k(0,1,0)$ or $\mathbf{k}=\mathbf{k}_{\parallel}=k(\cos{\theta},0,\sin{\theta})$ in Eq.(1) corresponds to scattering out-of-plane or in-plane respectively ($\theta$ represents an arbitrary in-plane direction),
\begin{equation}
\begin{gathered}
     S_{TD}(\mathbf{k}_{\perp},t)\propto\chi^{(n+d)}\rho(t)k^{-(n+d)}\\
     S_{TD}(\mathbf{k}_{\parallel},t)\propto\rho(t)k^{-(n+d)}
\end{gathered}
\end{equation}

These results show that $\eta$ is independent of the direction of $\mathbf{k}$. However the intensity is scaled by a factor of $\chi^{(n+d)}$, making the out-of-plane direction much more intense.

The in-plane intensity is clearly rescaled by $\chi^{-(n+d)}$ and compressed to lower wavevectors compared to the out-of-plane intensity. Nevertheless, the coarsening occurs in both directions.
In Fig. S3(a) we show the normalized intensity as a function of delay integrated over the red rectangle shown on the detector image in Fig. S3(b). The integration region is in the in-plane $a^*$ direction. The trace in Fig. S3(a) exhibits an initial suppression and later an increase of the normalized intensity that qualitatively resembles the black to yellow traces in main text Fig. 1(c), indicating that similar coarsening dynamics take place in-plane. However, this trace originates from wavevectors around $|\mathbf{k}|\approx0.0016$\AA$^{-1}$, much closer to the peak than those in Fig. 1(c), which are taken at $|\mathbf{k}|\approx0.02$\AA$^{-1}$$-0.06$\AA$^{-1}$. This shows that because of the substantial anisotropy in this system, $\chi$ is large, so the in-plane coarsening dynamics appears very close to the supperlattice peak. For these practical reasons we chose to focus our analysis along the $b$ direction where the intensity is stronger and distributed over the detector.

 \begin{figure}[ht]
\centering
\includegraphics[trim= 0 0 0 0,clip,width=0.95\linewidth]{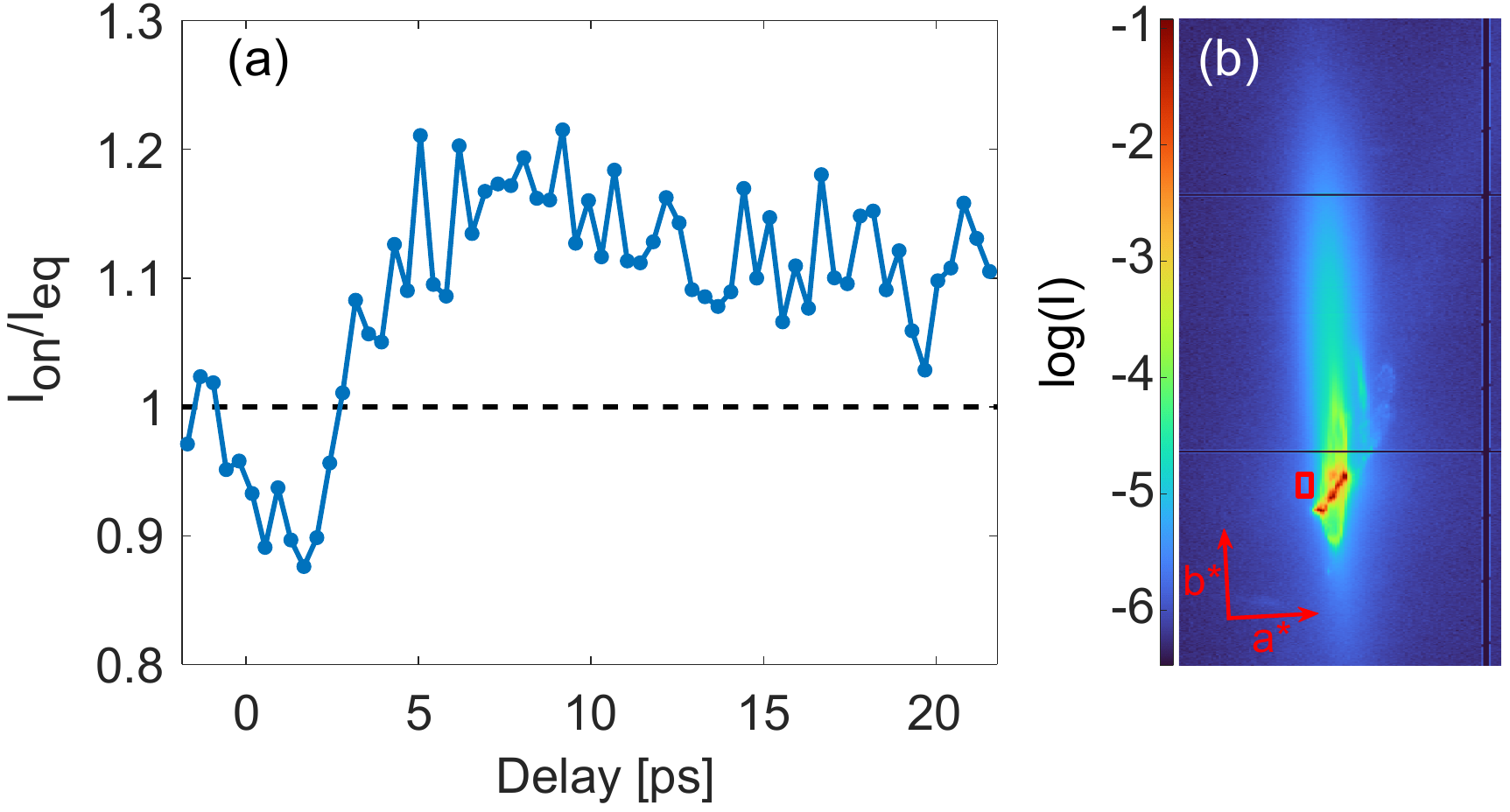}
\caption{{\bf Diffuse intensity in the in-plane direction.} (\textbf{A}) x-ray intensity versus $t$, normalized by the equilibrium intensity. The scattering was integrated in a region along the $a^*$ direction, marked by the red rectangle in (B). (\textbf{B}) Detector image of the $\left(2,2,1-q_{CDW}\right)$ superlattice peak, integrated over all pump-probe delays, showing the integration region for (A) (red rectangle).}
\end{figure}

\newpage
 \section*{Movie 1}
 {\bf Vortex Strings Dynamics.} Ginzburg-Landau simulation of the dynamics of the order parameter following a fast quench to the low symmetry free energy. The movie shows the surface $|\psi(x,t)|^2=0.5$ evolving in time, showcasing the formation, dynamics and annihilation of vortex strings.

\newpage
%